\documentclass[journal]{IEEEtran}

\usepackage{cite}
\usepackage{amsmath,amssymb,amsfonts}
\usepackage{graphicx}
\usepackage{textcomp}
\usepackage{xcolor}
\usepackage{mathtools}
\usepackage{algorithm}
\usepackage{algorithmicx}
\usepackage{algpseudocode}
\usepackage{amsmath}
\usepackage{times}
\usepackage{mathptmx}
\usepackage{multirow}
\usepackage{subfigure}
\usepackage{caption}
\usepackage{color,xcolor}
\graphicspath{{figure/}}
\def\BibTeX{{\rm B\kern-.05em{\sc i\kern-.025em b}\kern-.08em
    T\kern-.1667em\lower.7ex\hbox{E}\kern-.125emX}}

\newcommand{\ie}{\textit{i}.\textit{e}., }
\newcommand{\etc}{\textit{etc}}

\begin{document}


\title{Design and Implementation of Open LoRa for IoT}

\author{
	\IEEEauthorblockN{Qihao Zhou\textsuperscript{1}, Kan Zheng\textsuperscript{1}\IEEEauthorrefmark{1}, Lu Hou\textsuperscript{1}, Jinyu Xing\textsuperscript{1}, Rongtao Xu\textsuperscript{2}}\\
	\IEEEauthorblockA{\textsuperscript{1}\textit{Intelligent Computing and Communication (IC$^2$) Lab} \\
		\textit{Beijing University of Posts and Telecommunications \\ zqh@bupt.edu.cn \\}
		\textsuperscript{2}  \textit{State Key Laboratory of Rail Traffic Control and Safety \\ Beijing Jiaotong University \\}
		\IEEEauthorrefmark{1} Corresponding authors: zkan@bupt.edu.cn
	}
}
\maketitle

\begin{abstract}
	Long Range (LoRa) network is emerging as one of the most promising Low Power Wide Area (LPWA) networks, since it enables the energy-constraint devices distributed over wide areas to establish affordable connectivity. However, how to implement a cost-effective and flexible LoRa network is still an open challenge. This paper aims at exposing a feasible solution of design and implementation, allowing users to conveniently build a private LoRa network for various IoT applications. Firstly, several typical application scenarios of LoRa network are discussed. Then, the LoRa system architecture is presented with the functionality of each component. We address the hardware design and implementation of LoRa Gateway, which is the bridge between LoRa nodes and LoRa network server. Especially, the paper contributes by proposing an improved software architecture of LoRa network server whose source codes are open on GitHub. Under the architecture, LoRa network server is divided into four decoupled modules and uses the messaging system based on streaming data for the interaction between modules to guarantee scalability and flexibility. Finally, extensive experiments are conducted to evaluate the performance of LoRa networks in typical environments.
\end{abstract}

\begin{abstract}
LoRa, LPWA, IoT, Microservice, Open Source
\end{abstract}



\section{Introduction}

With the development of low power wide area (LPWA) technologies, the flourish of various Internet of Things (IoT) applications is changing the way of living in the world. In particular, it is estimated that LPWA devices will grow to 339 million in 2025 \cite{LoRa_Alliance_year_report}. Due to its unique technical features, Long Range (LoRa) network has attracted significant attention from industry and academic sectors in recent years. LoRa operates in the unlicensed Industrial Scientific and Medical (ISM) band and offers the long-distance connectivity to low-power devices. On the other hand, LoRaWAN is proposed by LoRa Alliance, which defines the network protocol in the medium access control (MAC) and network layers \cite{what_is_LoRaWAN}. A completed LoRa network primarily consists of several components, \ie LoRa Node, LoRa Gateway, LoRa network server and LoRa application server. It is worth mentioning that LoRa Nodes usually operate on battery power and have limited computational capabilities. LoRa network server is responsible to have sufficient ability to provide the packet processing services and support massive number of LoRa Nodes. Therefore, the architecture and implementation of the LoRa network server have a significant impact on the capacity of LoRa system.
 
There has been vast amounts of work being explored for LoRa network. Firstly, some of studies focused on comparing LoRa with other LPWA technologies such as Sigfox and NB-IoT \cite{Low_power_wide_area, LoRaWAN_review, survey_lpwa}. The authors in \cite{lora_scale, understanding_limits, performance_lora} analyzed and discussed the wide variation of performance in specific scenarios as well as the capabilities and limitations. {\color{black} These studies used analytical and simulation-based approaches to carry out the experiments and gave the processing performance of background server. There are various challenges when deploying LoRa network in practical scenario.} Moreover, several IoT cloud platforms have been proposed to process and manage IoT data \cite{data_management_platform, virtualization_in_IoT, SMDP}. However, LoRa Nodes are distributed over wide areas and utilize LoRa Gateway to establish the connection with Internet. These services provided by IoT cloud platforms are failed to be used directly by LoRa Nodes. A LoRa network server project is necessary to process the packets forwarded from/to LoRa Gateway. In \cite{LoRaCloud}, the authors proposed an implementation of LoRa network server on the OpenStack named as LoRaCloud without the performance evaluation. In addition, there are some projects available for building and managing LoRa network. {\color{black}For example, The Things Network (TTN) has begun to release an open source platform for LoRa network \cite{TTN}.} However, independent deployment of TTN-based LoRa network server is cumbersome, which is not conducive to agile development. {\color{black} Another popular open source project is named LoRaServer \cite{LoRaServer}. LoRaServer presented a ready-to-use solution including design and implementation. But the centralized architecture leads to excessive functionality concentrated on one module. It may have a negative impact on the throughput and scalability of entire system.} Therefore, an open LoRa network with high performance and flexibility is highly desirable.

In this paper, the typical application scenarios and requirements of LoRa network are presented firstly. Then, the LoRa system architecture is discussed as well as the functions of each component. In order to reduce the difficulty of deploying private LoRa networks, the hardware and software design are proposed in details. {\color{black} LoRa Gateway establishes the connection between massive LoRa Nodes and LoRa network server, so its function and performance are crucial. The paper focuses on the design and implementation of LoRa Gateway, which is able to support multi-mode communication and meets the needs of outdoor deployment.} Moreover, an improved LoRa network server architecture is proposed. The architecture follows the principles of microservice, allowing for maximum control and flexibility. As a result, the LoRa network server is divided into four modularized and low-coupling modules, \ie Connector, Central Server, Join Server and Network Controller. They are responsible for different functions such as protocol analysis or node activation individually. Various advanced techniques are used in the implementation of LoRa network server to guarantee high performance and scalability. {\color{black}Modular design allows processing modules to be deployed in separated processes or servers and help users flexibly extend custom functionality in the corresponding module.}

Aiming at providing insights about the design and implementation of LoRa system, the main contributions of this paper are summarized as follows:

\begin{enumerate}
	\item The design and implementation of LoRa Gateway are proposed as a feasible solution. 
	\item We propose an improved LoRa network server architecture and implement it with an open-source project on GitHub\footnote{ https://github.com/xisiot/lora-system}.
	\item Based on the proposed hardware platform and software architecture, a LoRa prototype system is deployed in urban environments and several experiments are conducted to evaluate the performance.
\end{enumerate}

The rest of this paper is organized as follows. Section \uppercase\expandafter{\romannumeral2} presents some typical application scenarios of LoRa networks. Section \uppercase\expandafter{\romannumeral3} describes LoRa network architecture. Section \uppercase\expandafter{\romannumeral4} gives the details on hardware design and implementation of LoRa Gateway. Section \uppercase\expandafter{\romannumeral5} proposes the improved architecture of LoRa network server. Section \uppercase\expandafter{\romannumeral6} demonstrates and analyses the experimental results. Finally, section \uppercase\expandafter{\romannumeral7} draws the conclusions.

\section{Typical Applications of LoRa Networks}
\begin{figure} [t]
	\centering
	\includegraphics[width=0.50\textwidth]{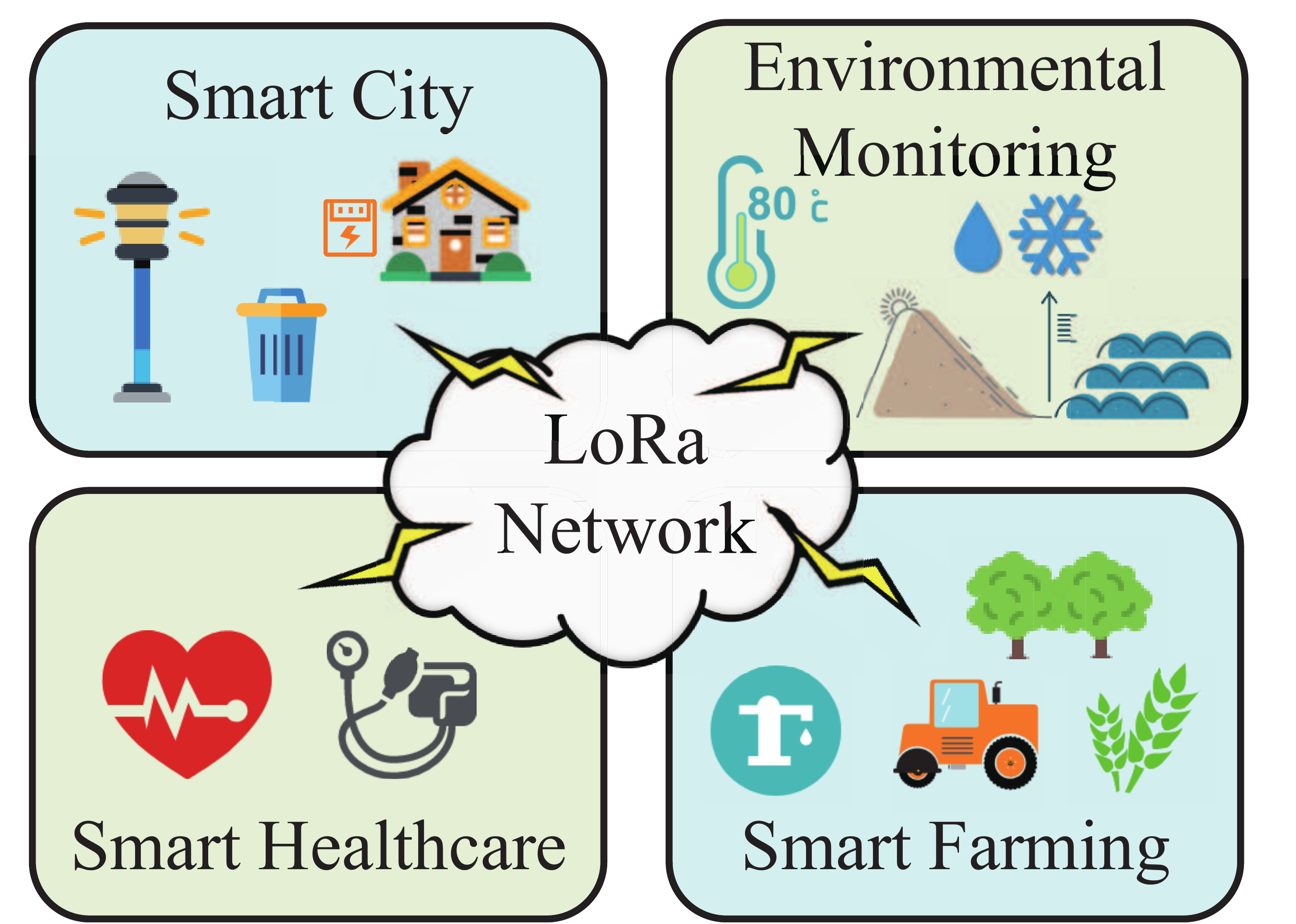}
	\caption{Illustration of typical application scenarios of LoRa networks.}
	\label{LoRa_Application}
\end{figure}

With the communication range of over ten kilometers and long battery lifetime, LoRa network is promising for the long range, low power and low cost IoT applications. However, in order to achieve these excellent characteristics, LoRa technology comes at the expense of low data rate \cite{lpwa_overview}. In this section, several typical applications and scenarios of LoRa network are discussed.

\subsection{Smart City}

LoRa network provides a reliable and feasible solution for various IoT applications in smart city, such as smart metering and waste management. More intelligently, these LoRa Nodes can automatically report sensing data and be controlled remotely through LoRa network. By connecting city services, these applications offer the opportunities to improve the efficiency of municipal operations and reduce maintenance costs.

\subsection{Environmental Monitoring}

Environmental monitoring is crucial for our society to provide real-time environmental indicators, such as air quality, temperature and humidity, and alert to critical situations, such as natural disasters and environment pollution. Taking into account both financial and energy requirements, it is suitable to implement a network with LoRa technology across large geographical areas. The low-power LoRa Nodes with various sensors are able to interact with the environment and report data opportunely to detect issues before becoming crises.

\subsection{Smart Healthcare}
Using information and communication technology in healthcare can lead to better treatment and disease surveillance for patients. However, the vast cost burden is a challenge for the healthcare industry. Low cost and reliable performance of LoRa network make it suitable for typical smart healthcare applications. Various biological information is collected by special on-body sensors and can be checked in time. All abnormal indicators are transmitted to health service providers or health professionals through LoRa network for the early detection and prevention of diseases.

\subsection{Smart Farming}
Smart farming is the application of modern technology into agriculture, aiming at improving the quantity and quality of agricultural products while reducing environmental impact and preserving resources. LoRa-based smart farming use cases, e.g., soil moisture monitoring and autonomous irrigation, have a great potential to deliver a more sustainable agricultural production. Long-range, low-cost and low-power features of LoRa technology enable the use of sensors to transfer information from the farm to the cloud where it can make more efficient operations and management.

\section{LoRa Network Architecture}

\begin{figure}[t]
	\centering
	\includegraphics[width=0.49\textwidth]{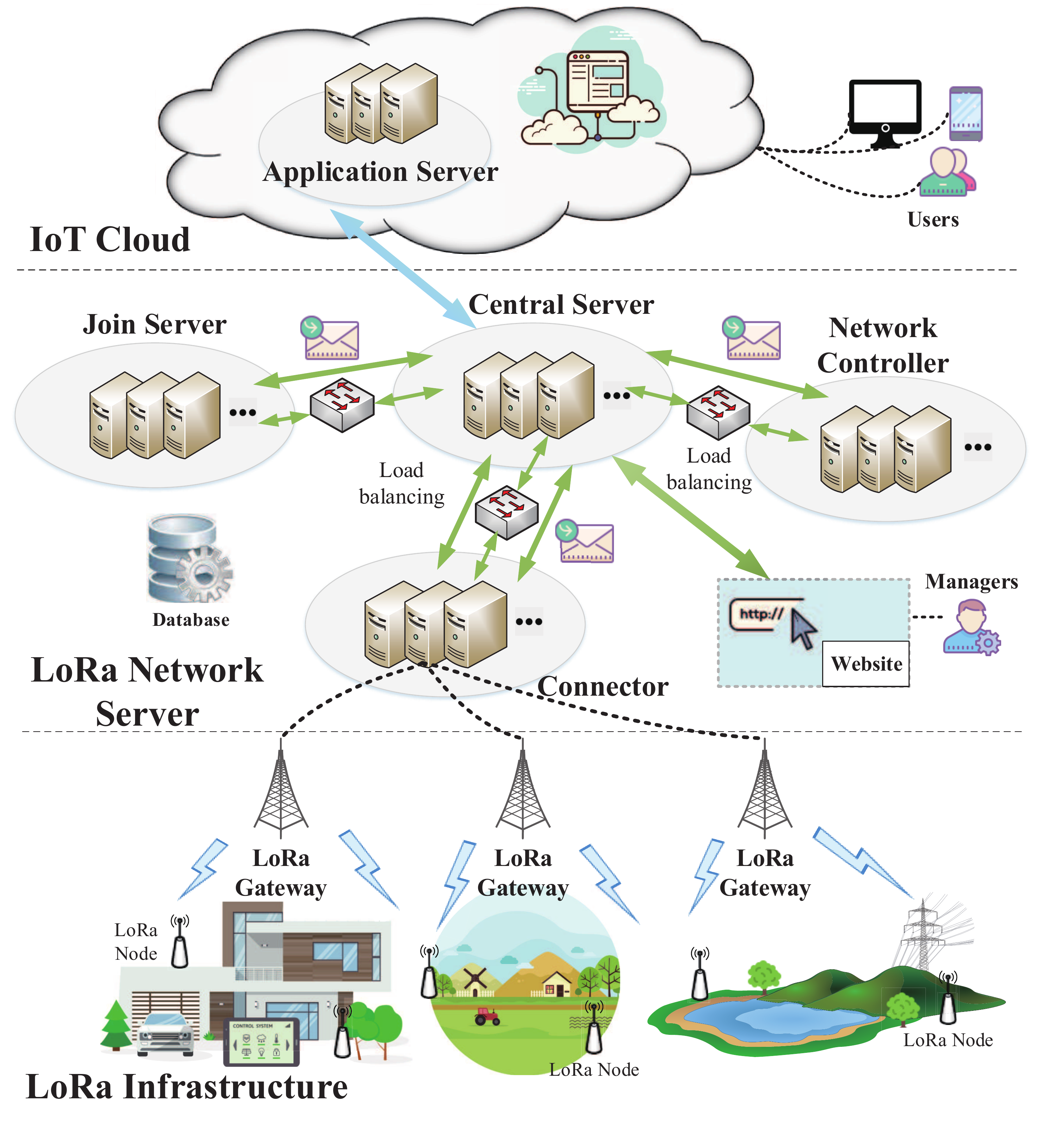}
	\caption{Illustration of LoRa network architecture.}
	\label{LoRa_Architecture}
\end{figure}

As illustrated in Fig.\ref{LoRa_Architecture}, a three-tier hierarchical LoRa network architecture is proposed. The functionalities of each component is described in this section.

\subsection{LoRa Node and Gateway}\label{3A}

LoRa Node and Gateway are fundamental components of the entire LoRa networks since it can interact with the environment as well as forwarding data to LoRa network server for analysis. LoRa Node mainly includes sensors, actuators, battery, \etc., and performs one or more tasks according to different application requirements. LoRa Gateway forwards all uplink radio packets to LoRa network server after adding metadata such as Signal-Noise-Ratio (SNR) and Received-Signal-Strength-Indicator (RSSI). Conversely, on the downlink, LoRa Gateway performs transmission requests coming from LoRa network server. Typically, it only transmits the received uplink/downlink packets as a relay without any interpretation of payloads. The hardware implementation details are described in the next section.

\subsection{LoRa Network Server}\label{3B}
LoRa network server receives packets forwarded from LoRa Gateway, and is responsible for packet processing and protocol analysis. {\color{black} Vast amount of functions need to be implemented on LoRa network server. For examples, checking the legality of end-devices, filtering duplicate uplink packets, performing network management mechanisms, scheduling acknowledgements and forwarding application layer data to the application server, \etc. A well-designed architecture of LoRa network server is important for efficient processing and convenient management. Especially, the flexibility allows users to build various applications on existing LoRa hardware devices and the scalability can meet the needs of supporting massive LoRa Nodes. In addition, a management framework is necessary to help users to register, manage and monitor their LoRa devices including nodes and gateways. Aiming at providing insight about the improvement of LoRa network server, the proposed design and implementation of LoRa network server are presented in Section \uppercase\expandafter{\romannumeral5}.}

\subsection{Application Server in IoT cloud}
LoRa Application Server is responsible for encryption, decryption and processing of application layer payloads. It can support various applications with different encryption or encoding methods, such as Protocol Buffer, to ensure data security and improve transmission efficiency. IoT cloud provides essential services through Application Program Interfaces (APIs) and then applications are provided to users. LoRa Application Server acts as a bridge between IoT cloud and LoRa network server so that users can control LoRa Nodes and enjoy applications of LoRa network through web browsers or smart phones anywhere. A detailed explanation of IoT cloud can be obtained in \cite{iotcloud}.

\section{Hardware Design and Implementation}

\begin{figure}[t]
	\centering
	\includegraphics[width=0.48\textwidth]{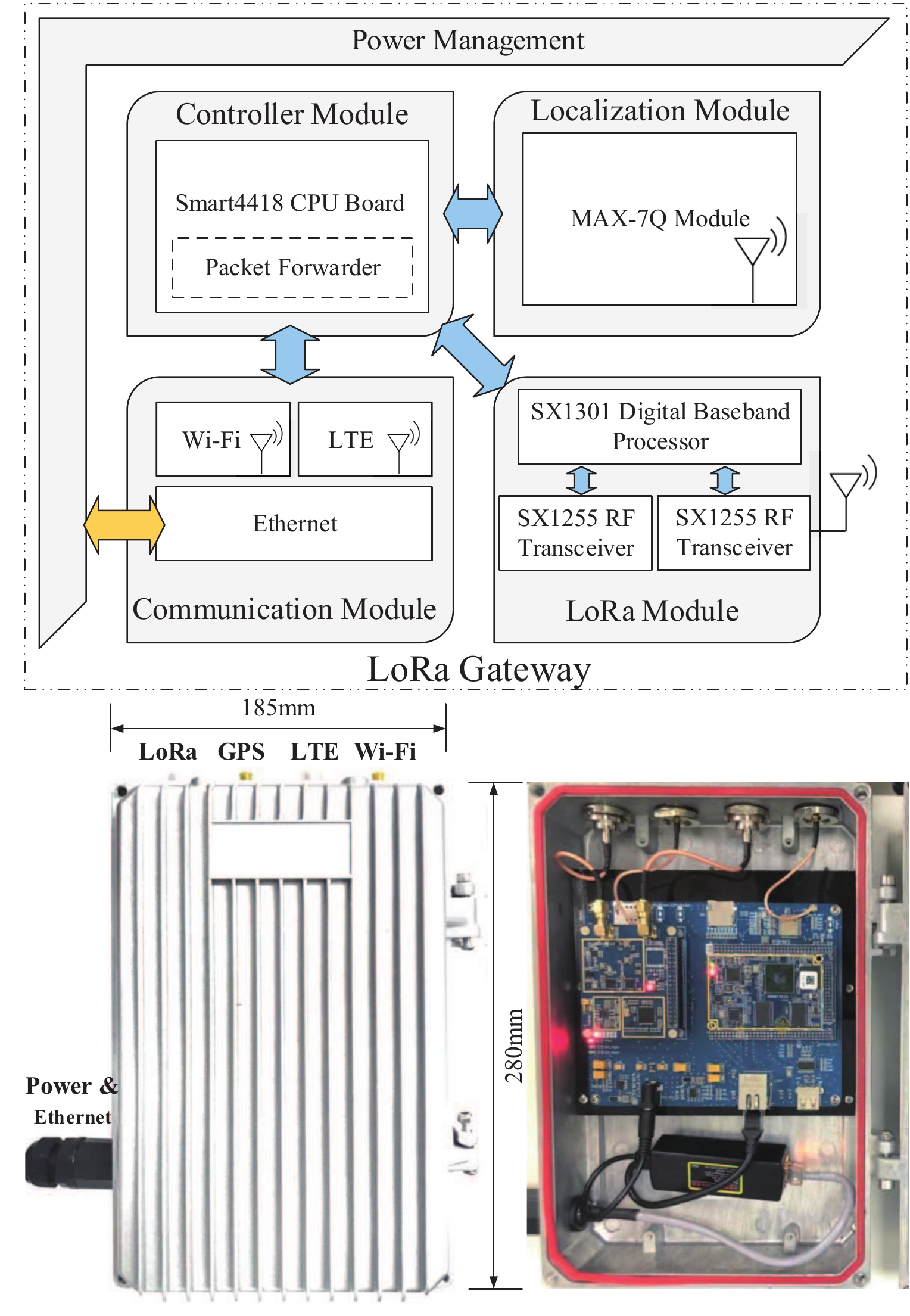}
	\caption{Illustration of LoRa Gateway and its modules.}
	\label{Gateway_Structure}
\end{figure}

The hardware of LoRa system includes LoRa Node and LoRa Gateway. The design of LoRa Node is dependent on its specific application requirements and one typical example can be found in our published work \cite{irrigation_system}. {\color{black}In this paper, we only focus on the design and implementation of LoRa Gateway, which acts as a relay between LoRa Node and LoRa network server. Due to LoRa technology’s long range capability, the LoRa Gateway is expected to cover dozens of square kilometers in the given environment. Therefore, it is necessary for LoRa Gateway to support two-way multi-channel communication based LoRa technology and have high signal receiving sensibility. Rich interfaces and high processing performance are the guarantee of scalability and large-scale access. In addition, LoRa Gateway should have good water-proof ability to support outdoor deployment.}

As shown in Fig.\ref{Gateway_Structure}, the LoRa Gateway consists of five functional modules, \ie controller module, localization module, power management module, LoRa module and communication module. Its degrees of protection provided by enclosures can meet the requirements of IP65 level. The implementation details are given as follows, \ie

\subsection{Controller Module}
All control functions are implemented in controller module. {\color{black}The Smart4418 board is chosen due to its good performance, which uses a quad core Cortex A9 Computer Processing Unit (CPU) with dynamic frequency scaling up to 1.4 $GHz$ and has 1 $GB$ Random Access Memory (RAM) and 8 $GB$ Embedded Multi Media Card (eMMC). Moreover, the board offers rich and advanced communication interfaces, such as up to 24 General-Purpose Input/Output (GPIO), 2 Serial Peripheral Interface (SPI) and Gbps Ethernet, \etc. These features make Smart4418 suitable for implementation of LoRa Gateway.}

In order to build a friendly runtime environment for software, the Debian system based on the Linux kernel is used in controller module. The packet forwarder program processes the communication protocol between LoRa Gateway and LoRa network server \cite{packet_forwarder}. Based on this program, LoRa Gateway has the ability to connect with the server using specific packets through a User Datagram Protocol (UDP) link.

\subsection{LoRa Module}
{\color{black}In order to achieve two-way multi-channel communication and excellent receiver sensitivity, the LoRa communication module consists of two SX1255 transceivers and one SX1301 digital baseband processor.} SX1255 is an integrated RF transceiver and is designed to operate over the 400-510 $MHz$ frequency bands. After SX1255 transceivers capture the RF signal, the data is concentrated into the SX1301 baseband processor via SPI. SX1301 contains 8 programmable parallel demodulation paths for LoRa channel and is able to demodulate simultaneously up to 8 packets. In addition, data-rate adaptation (ADR) scheme is supported by our designed LoRa Gateway. LoRa Node is able to transmit the packets with different data rate on a given channel as required.

\subsection{Communication Module}
The LoRa Gateway has the other communication modules including Wi-Fi, LTE and Ethernet. Depending on the deployment environment, the reliable connection is able to be established from gateway to LoRa network server through one of communication mode.

\subsection{Localization Module}
{\color{black}Localization module uses the U-blox’s MAX-7Q for localization and time synchronization. The module can provide the position accuracy of less than 5 meters with low power and short search time from Global Positioning System (GPS).} The GPS related information is transmitted to controller module through the serial port and then is parsed according to standard protocols.

\subsection{Power Management Module}
Power Over Ethernet (POE) is adopted to simplify the deployment of LoRa Gateway. The power management module uses TPS54332D to offer 3.3 $V$ and 5 $V$ voltages for different components.

\section{LoRa Network Server Design and Implementation}
{\color{black} Aiming at providing an open flexible solution for both allowing users to conveniently deploying the LoRa network and supporting massive LoRa Nodes, we propose the improved architecture of LoRa network server and discuss its details as well as the implementation in this section.}

{\color{black}
\subsection{Related Works on Architecture Design}
 \begin{figure}[]
	\centering
	\subfigure[]{
		\begin{minipage}[a]{0.48\textwidth}
			\includegraphics[width=1\textwidth]{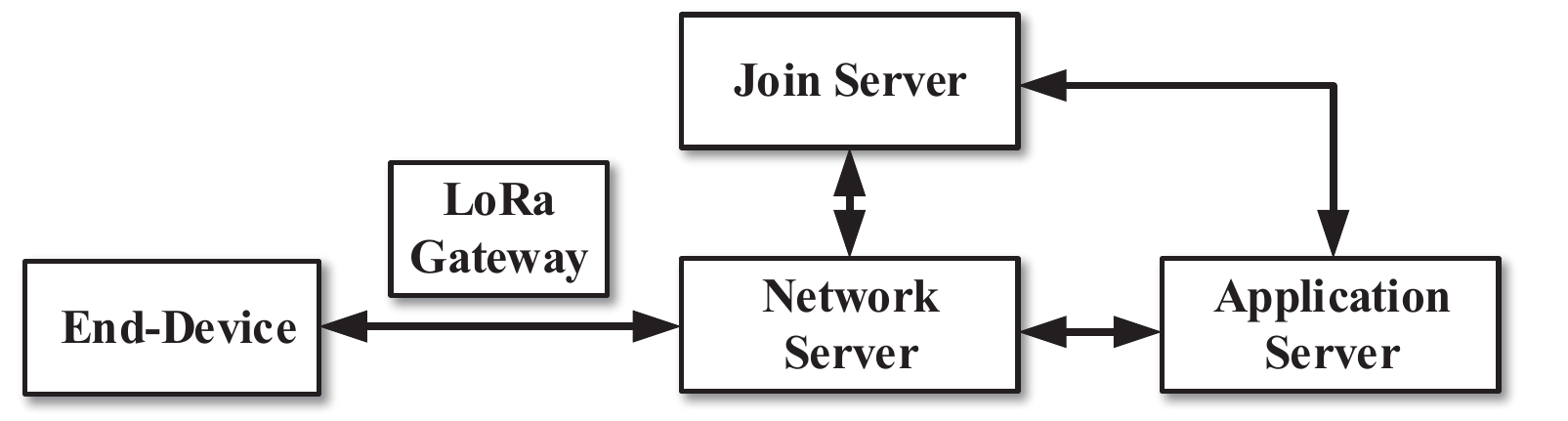} \\
		\end{minipage}
	}
	\subfigure[]{
		\begin{minipage}[b]{0.48\textwidth}
			\includegraphics[width=1\textwidth]{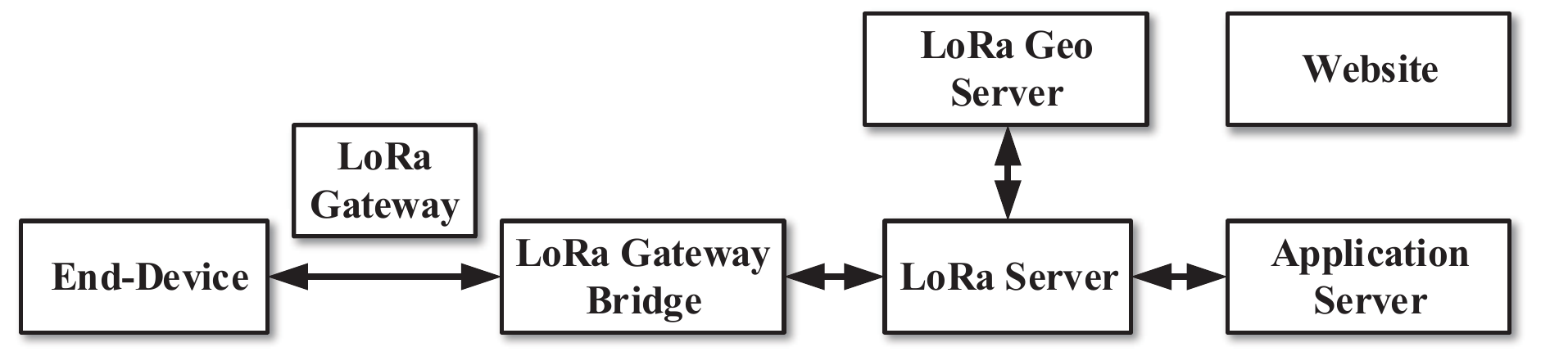} \\
		\end{minipage}
	}
	\caption{(a) The LoRa network architecture given by LoRaWAN. (b) The LoRa network architecture proposed by CableLabs \cite{LoRaServer}.}
	\label{compare_LoRa}
	\vspace{-1em}
\end{figure}

In the paper, we describe two typical architecture designs for comparison, i.e., the LoRa network architecture given by LoRaWAN and proposed by CableLabs \cite{LoRaServer}, which are shown in Fig.\ref{compare_LoRa} (a) and (b) respectively.

In the reference architecture given by LoRaWAN, Network Server is the center of star topology and is expected to implement  all the main functions. However, it may become a potential bottleneck because integrating excessive functionality requires more capabilities \cite{Challenges}. It also lacks of expansibility and cannot resist single point of failure. In addition, LoRaWAN only introduces the typical architecture without giving any concrete implementation. Different realizations may have a significant impact on the performance of LoRa network server.

As shown in Fig.\ref{compare_LoRa} (b), the architecture proposed by CableLabs separates two modules from LoRa Server, \ie LoRa Gateway Bridge and LoRa Geo Server \cite{LoRaServer}. They are designed to establish communication with LoRa Gateway and provide geolocation services respectively. Registration-related features such as handling join requests are integrated into LoRa Server. Compared with our proposed architecture, there are two main differences in the design concept. Firstly, the interaction between modules is implemented by gRPC while the messaging system is used in our proposed solution. The advantages of using the messaging system are elaborated later. On the other hand, similar to the architecture proposed by LoRaWAN, the centralized architecture leads to excessive functionality concentrated on one module, \ie LoRa Server. We propose a LoRa network server architecture that separates functionality into four flexible modules to attenuate the negative impact of centralization.
}
\subsection{Design of LoRa Network Server}

\begin{figure} []
	\includegraphics[width=0.50\textwidth]{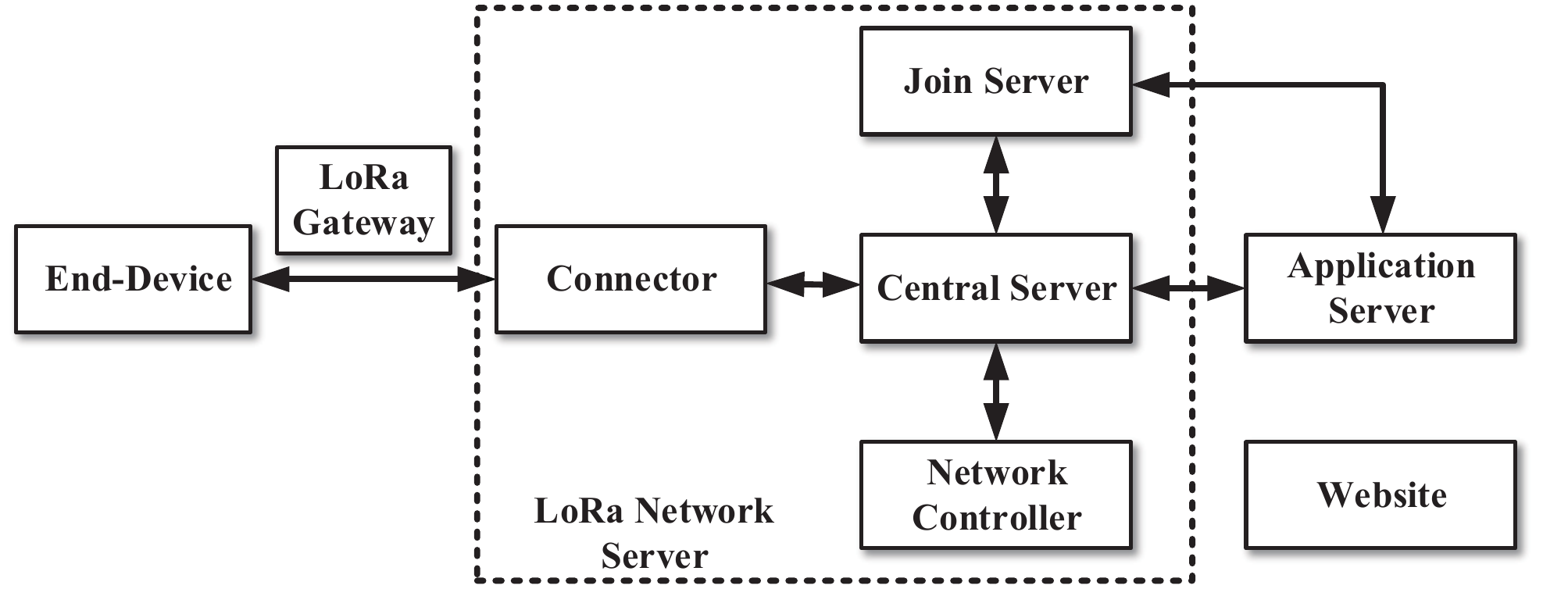}
	\caption{The improved architecture of LoRa network server.}
	\label{LoRa_Architecture_Simple}
	\vspace{-1em}
\end{figure}

As illustrated in Fig.\ref{LoRa_Architecture_Simple}, the improved architecture of LoRa network server is divided into four modules, \ie, Connector, Central Server, Join Server and Network Controller. It is worth mentioning that Connector performs packet parsing and discards illegal packets directly, which both improves security and reduces the occupation of computing resources. Network Controller is also separated from the traditional architecture in order to flexibly implement and perform network management schemes, such as Adaptive Data Rate (ADR) scheme.

Modularizing and splitting main functions into different modules can distribute the computational load as well as reducing performance requirements of the computer server, such as CPU and memory. The division of functions is in line with the principle of microservices in order to ensure low coupling between modules. Therefore, these modules can be deployed flexibly in separated processes or servers, allowing for agile deployment. All the modules in LoRa network server also can be quickly deployed by Docker containerization \cite{docker}. We package all the necessary components into a completed file system including project codes, databases and libraries. The method simplifies deployment steps as well as guaranteeing that each module runs correctly regardless of its environment. The key points in the architecture are elaborated as follows.

\subsubsection{Interaction Between Modules}
\begin{figure} [b]
	\includegraphics[width=0.50\textwidth]{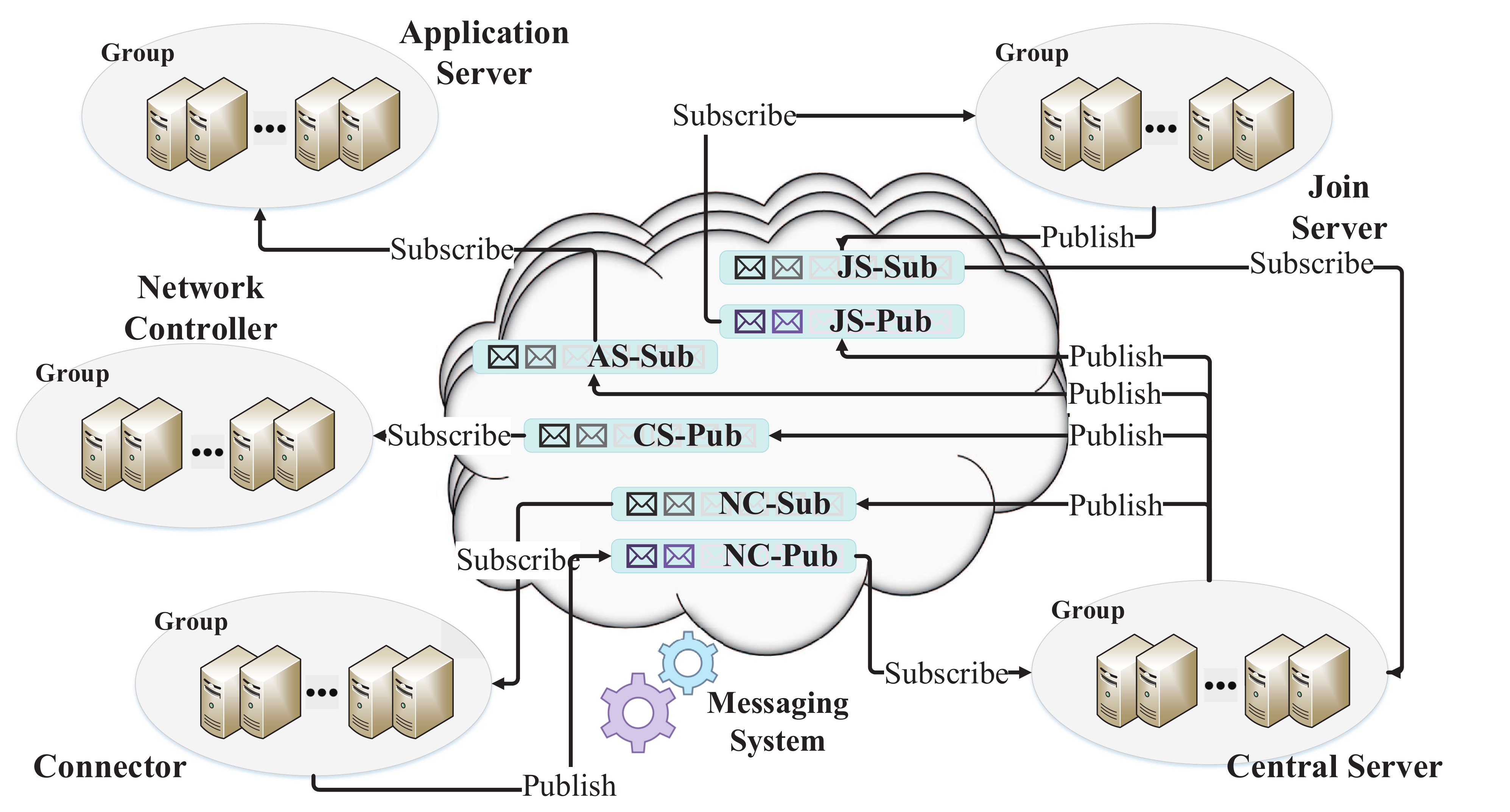}
	\caption{The design of interaction between modules.}
	\label{Interaction_Models}
\end{figure}

The interaction between modules is critical for the overall performance of LoRa network server. The publish-subscribe messaging system, \ie Apache Kafka, is used to establish reliable communication between models. As shown in Fig.\ref{Interaction_Models}, the topic is a name of streaming data pipeline in messaging system and different topics are designed for each module according to requirements. Each module subscribes one or more topics to fetch the corresponding data as a consumer and publishes messages to specific topics as a producer. The streaming data pipelines are acting as an abstract layer between consumer and producer. As long as the message format is specified, producers can be "ignorant" of the consumers. {\color{black} Moreover, the messaging system is able to provide both ordering guarantees and reliable asynchronous processing to address high concurrency challenges.}

\subsubsection{Load balancing}
Load balancing is an essential design to handle a massive number of concurrent packets. In the messaging system, the consumer group has the concept of clustering the same modules for load balancing and fail-over. It is worth mentioning that each message published from the producer is assigned to one consumer in the subscribing consumer group. The scheduler operates the round-robin scheduling policy for all available consumers. Load balancing improves overall throughput and avoids bottlenecks caused by excessive load on a single processing module. {\color{black}In addition, load balancing also enhances the fault tolerance of LoRa network server. When a given single module fails, the data which needs to be processed can be immediately switched to other parallel modules without service interruption.}

\subsubsection{Database}
In order to meet the storage requirements of heterogeneous data, Structured Query Language (SQL) and NoSQL databases are selectively used in LoRa network server. The persistent data with definite relationships, such as client intrinsic information, are stored in MySQL database. However, the main bottleneck of SQL database is the inefficiency in handling high concurrent read and write operations \cite{sql_nosql}. Consequently, NoSQL databases are able to provide real-time and high-efficiency services for data storage. Therefore, all session-related and non-persistent data, such as logs and duplicate packets, is stored by NoSQL databases. In addition, an efficient caching mechanism is necessary to ensure low latency and high performance when dealing with large amounts of concurrent packets.

\subsection{Implementation of LoRa Network Server}
The implementation of LoRa network server are described with the functionality of each module. Aim at efficiently processing large amounts of packets,  different types of content are assigned to different modules for processing. These modules are built on microservices principles with the goal of decoupling their duties and achieving distributed deployment. Moreover, a website project is developed as a management framework, which allows operators to register devices, monitor devices and retrieve historical data, \etc.

\subsubsection{Connector}
Connector is the bridge between LoRa Gateway and Central Server, which provides the services for parsing and packaging the payload based on LoRaWAN specification \cite{LoRaWAN_specification}. LoRa Gateway connects with a given Connector through an IP back-bone and uses UDP for packet transmission. Firstly, Connector performs legality verification on LoRa Nodes and Gateway to avoid unauthorized access. For example, if LoRa Gateway has been legally registered, Connector immediately replies an acknowledgment to indicates that the packet is received. Otherwise, the packet is discarded directly. Moreover, parsing and packaging of a packet require many necessary database operations. Separating the function from traditional network server performs well to spread the processing pressure and improve the utilization of all accessible resources. On the other hand, packets associated with LoRa node activation process are ignored and forwarded directly to Central Server.

\subsubsection{Central Server}
Central Server is responsible for data management and service scheduling. Different modules are scheduled by Central Server according to the requirements of data processing. Depending on the type of uplink packet, the information in packet is separated into the specific formats in order to facilitate subsequent processing. Then, the original application data is fed into Application Server, the MAC commands are sent to Network Controller and the join packets are forwarded to Join Server. In addition, the transmitted data is collected and stored by Central Server. Managers can use web browsers to check up the uplink/downlink packets and monitor the running states of LoRa node and Gateway in real time.

Since LoRa Nodes are not associated with a specific LoRa Gateway, the single packet may be received by multiple gateways simultaneously. To avoid the waste of radio resources due to redundancy, Central Server is essential for filtering duplicate packets. Only one of duplicate packets is fed into the subsequent processing module. Moreover, the responsibility of Central Server includes scheduling of downlink packets. The transmission information such as RSSI and SNR attached in the duplicate packets is not discarded and then one optimal gateway is selected to send the downlink packet through exploiting this information. Finally, Central Server determines the downlink parameters according to related configurations and reads the downlink payload from two data queues which are responsible for application data and MAC commands.

\subsubsection{Join Server}
The services provided by Join Server include handling activation requests and generating session keys. There are two methods to activate LoRa Nodes, \ie Over-The-Air Activation (OTAA) and Activation By Personalization (ABP). For OTAA method, LoRa Node must follow a join procedure consisting of the join request and the join accept exchange. To secure the activation process, the join request packet is only parsed by Join Server and then the session keys are derived from both the pre-shared key and the transmitted random number. Finally, the join accept packet is encapsulated and encrypted by the corresponding key in Join Server. Furthermore, LoRa Nodes using ABP method need to be registered on the website through specific management interfaces.

\subsubsection{Network Controller}
\begin{figure} [!t]
	\includegraphics[width=0.48\textwidth]{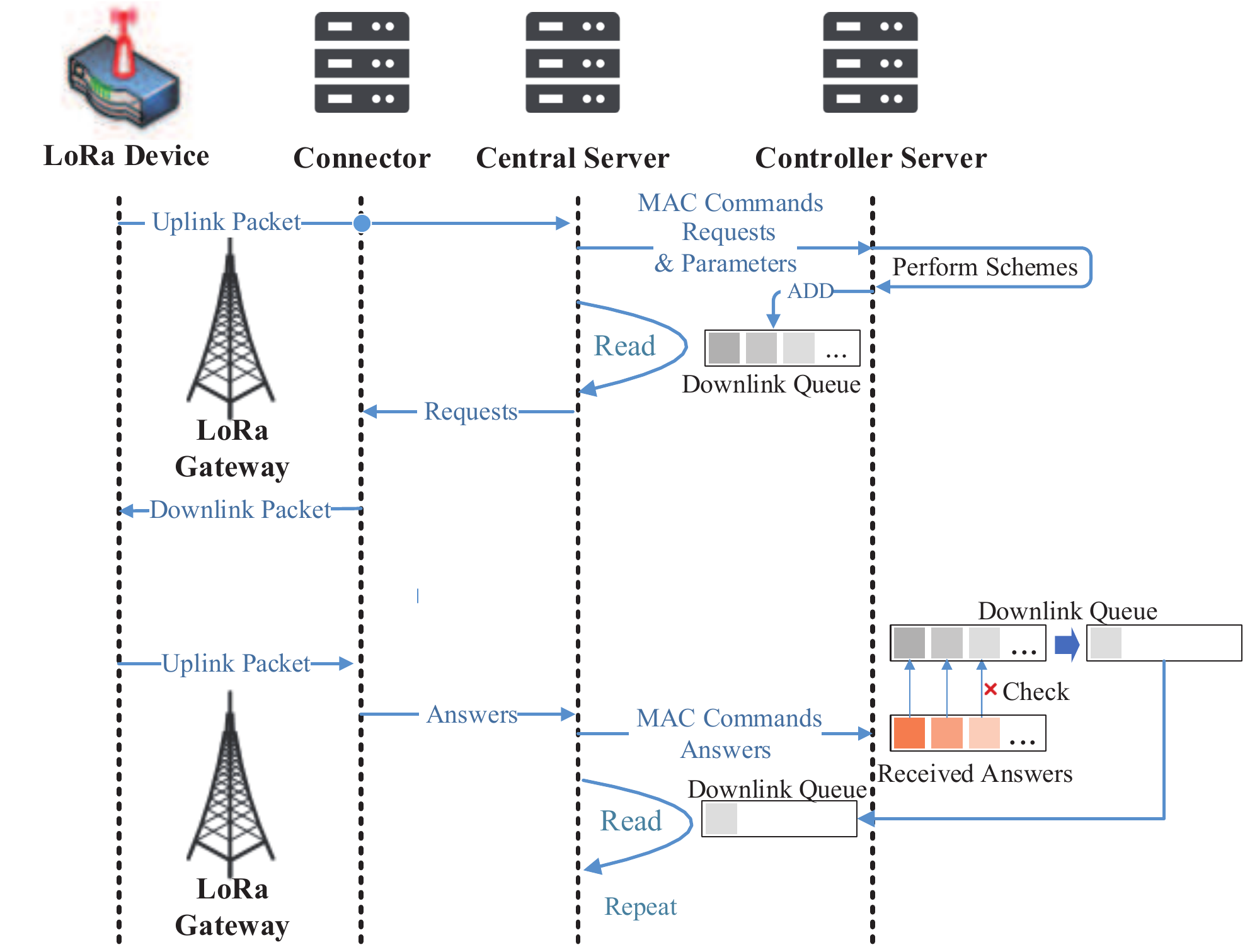}
	\caption{Illustration of the workflow of Network Controller.}
	\label{NC_MacCommands}
\end{figure}
The functions of Network Controller focus on processing and managing MAC commands. MAC commands are exchanged between Network Controller and LoRa Nodes in order to modify associated configurations or adjust transmission parameters \cite{LoRaWAN_specification}. Network Controller is stripped from classic LoRa network server, which flexibly performs Mac commands and new network management schemes. The well-designed schemes such as Adaptive Data Rate (ADR) are implemented since they can improve both reliability and capability of LoRa networks \cite{Energy-Efficient}. Fig.\ref{NC_MacCommands} shows the workflow of Network Controller. Take the implementation of ADR scheme as an example, the transmission parameters of the uplink packets are conveyed from Central Server. Then Network Controller performs the ADR scheme and determines whether to adjust either the data rate or transmission power. Finally, the MAC commands are added to the downlink queue and sent by the downlink packet. When receiving the acknowledgement packet, the successful request commands are removed from the queue. Otherwise, the failed commands are reserved for the next transmission.

\section{Experimental Results and Analysis}
In order to evaluate the performance of proposed LoRa network, a prototype LoRa network has been implemented and deployed in typical urban environments. Not only the coverage but also the network server performance of LoRa system are evaluated in this section.

\subsection{Coverage performance and analysis}
{\color{black} The LoRa Gateway with whip antenna is vertically installed on the roof of a 15-floor building at the center of our campus. The omnidirectional antenna has a gain of 10 $dBi$ and an operating frequency range from 400 $MHz$ to 470 $MHz$.} Then, the packets are collected through LoRa transmission and then forwarded to LoRa network server by LTE \cite{irrigation_system}. Both the LoRa Gateway and Nodes operate at the frequency band of 433 $MHz$. In order to evaluate the maximum distance of the LoRa transmission, LoRa Nodes adopt the maximum spreading factor, \ie SF=12, to transmit the packets. The main parameters in field trials are given in Table \ref{node_parameters}.

\begin{table}[]
    \caption{LoRa Node Parameters}
    \label{node_parameters}
    \begin{center}
    \begin{tabular}{l|l}
    \hline
    \textbf{Parameters} & \textbf{Value} \\ \hline
    Carrier frequency & 433 $MHz$  \\ \hline
    Preamble          & 8 $Symbol$ \\ \hline
    Transmit power    & 20 $dBm$   \\ \hline
    Coding rate       & 4/5      \\ \hline
    Spreading factor  & 12       \\ \hline
    Bandwidth         & 125 $kHz$  \\ \hline
    \end{tabular}
    \end{center}
\end{table}
\begin{figure}[!t]
	\vspace{-2em}
	\centering
	\includegraphics[width=0.50\textwidth]{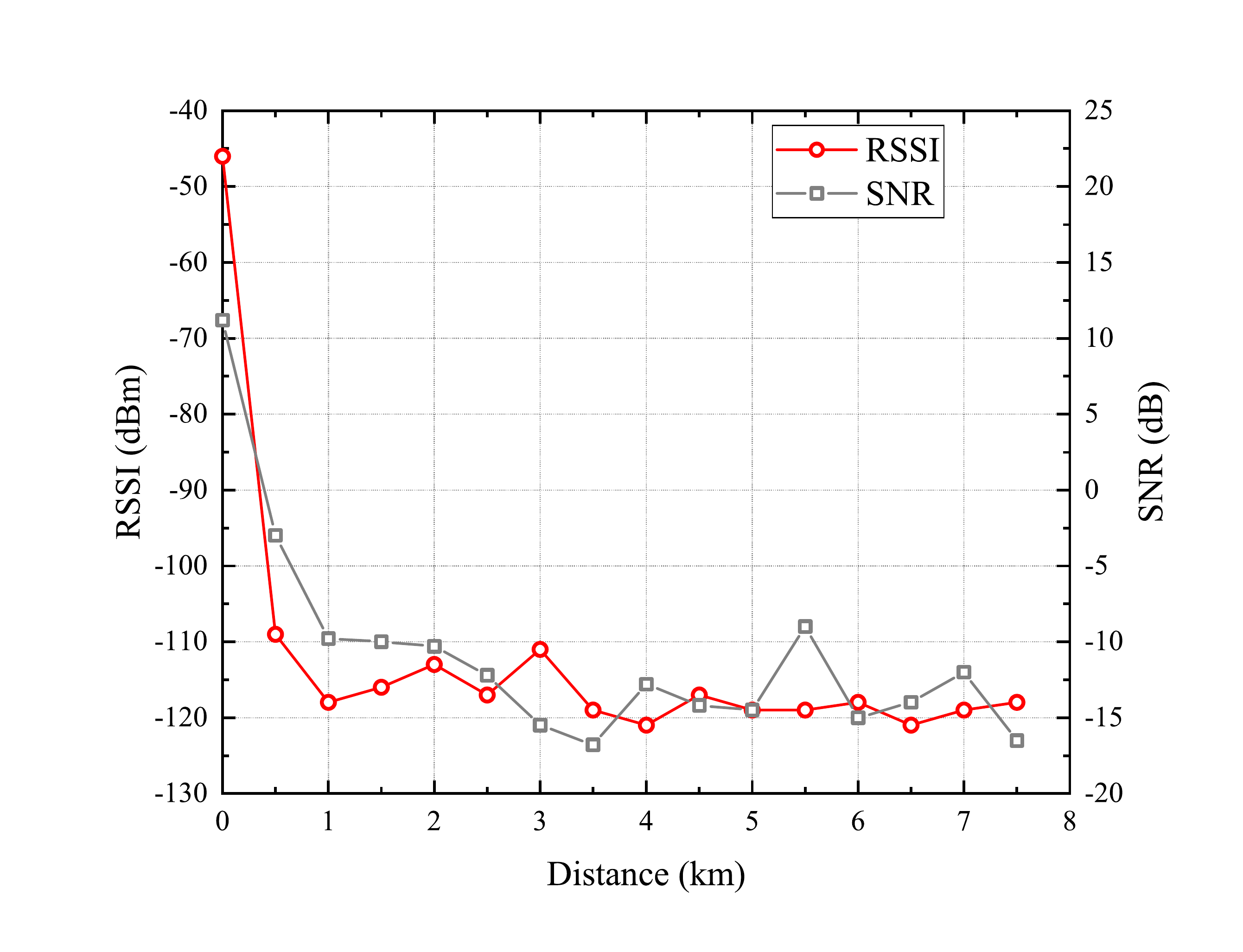}
	\caption{Coverage performance of LoRa transmission.}
	\label{snr}	
	\vspace{-1em}
\end{figure}
In the experiments, several LoRa Nodes continuously send uplink packets and gradually move away from the location of LoRa Gateway. The received signal quality parameters are measured about every 0.5 $km$. As shown in Fig.\ref{snr}, the RSSI and SNR decrease sharply within 1 $km$, while vary slowly with the increase of distance. It can be seen that the maximum distance that LoRa Gateway can receive the uplink packets is about 7.5 $km$, while the minimal RSSI and SNR is -118 $dBm$ and -16.5 $dB$ respectively.

\subsection{LoRa Network server performance and analysis}
In order to evaluate the network performance of LoRa system, the proposed modules and the stress testing program are deployed on two separated VMs. The main parameters of the VMs are shown in Table \ref{vm_arameters}. The testing program is developed based on Locust, which can emulate thousands of concurrent users on a single VM \cite{locust}. In our experiments, each LoRa node is treated as a user and generates an uplink application packet about every 40 seconds. We assume that all the packets are successfully received by LoRa Gateway since LoRa Gateway has multi-channel communication capability and powerful processing performance. The testing program is designed to directly send/receive the uplink/downlink packets by UDP to/from the network server \cite{packet_forwarder}. The number of LoRa Nodes is set from 400 to 14,000 as required. Each uplink packet should be replied and the response packets are received by LoRa Nodes if the network server handles the packets successfully. Otherwise, if no response is received within 5 seconds, the processing of uplink packet is regarded as a failure. When the number of LoRa Nodes increases, the throughput under the given situation is measured. Meanwhile, the median response time and average CPU utilization of Connector and Central Server are also used as reference indicators. Our experiments are mainly focused on Connector and Central Server since both of them have significant impacts on the quality of services under consideration.

\begin{table}[]
    \caption{Virtual Machine Parameters}
    \label{vm_arameters}
    \begin{center}
    \begin{tabular}{l|l|l}
    \hline
    \textbf{Parameters}   & \textbf{Server VM}                                                      & \textbf{Test VM}                                                          \\ \hline
    CPU                   & \begin{tabular}[c]{@{}l@{}}Intel(R) Xeon(R) \\ CPU E5-2680\end{tabular} & \begin{tabular}[c]{@{}l@{}}Intel(R) Xeon(R) \\ CPU E5-2609 v4\end{tabular} \\ \hline
    RAM                   & 8 $GB$                                                                    & 16 $GB$                                                                     \\ \hline
    Number of logical CPU & 5                                                                       & 8                                                                         \\ \hline
    Operating System      & Ubuntu 14.04.3 LTS                                                      & Ubuntu 16.04.3 LTS                                                        \\ \hline
    \end{tabular}
    \end{center}
\end{table}
\begin{figure}[!t]
	\vspace{-2em}
	\centering
	\includegraphics[width=0.53\textwidth]{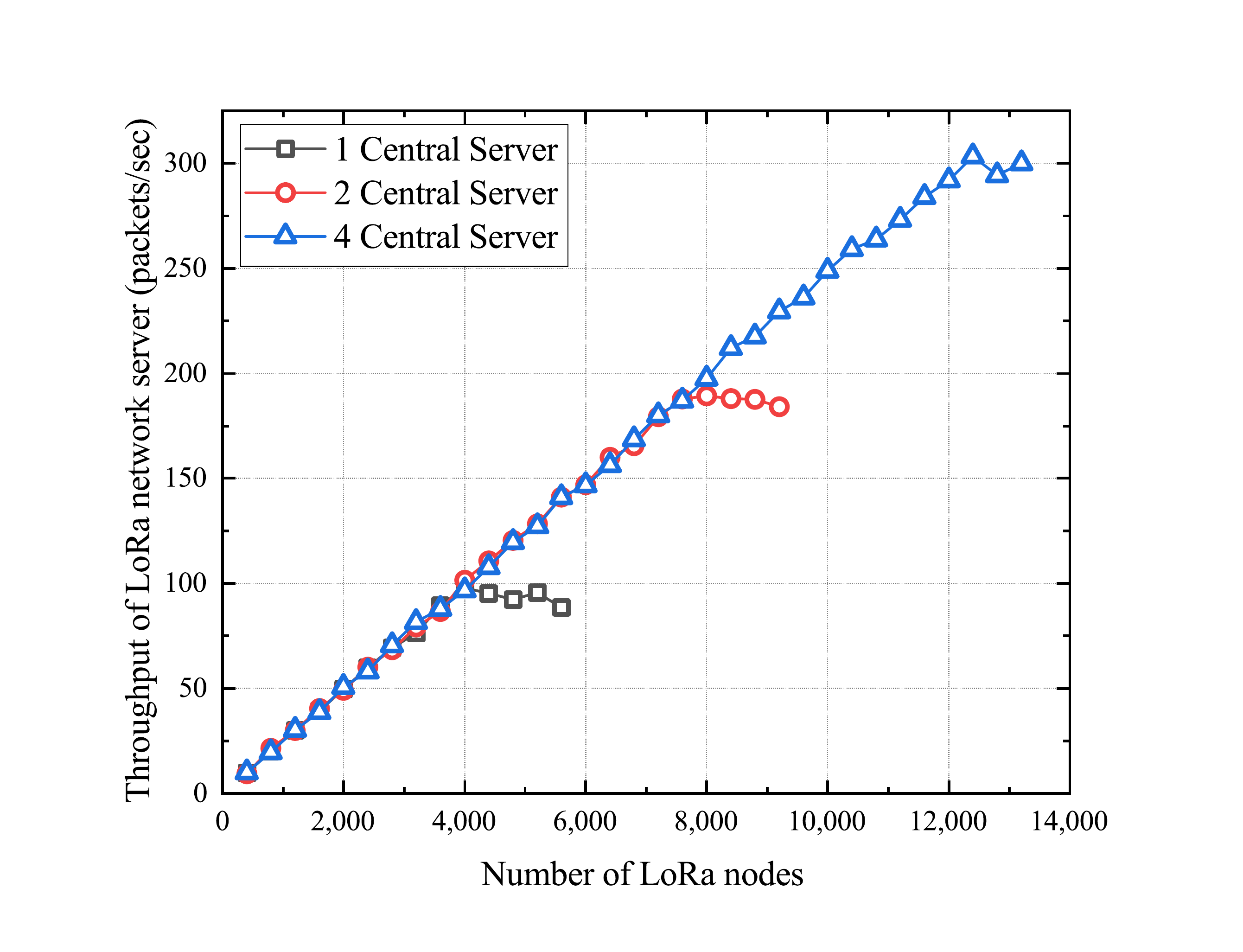}
	\caption{Throughput of LoRa network server.}
	\label{throughput}	
\end{figure}
\begin{figure}[!t]
	\vspace{-1em}
	\centering
	\includegraphics[width=0.53\textwidth]{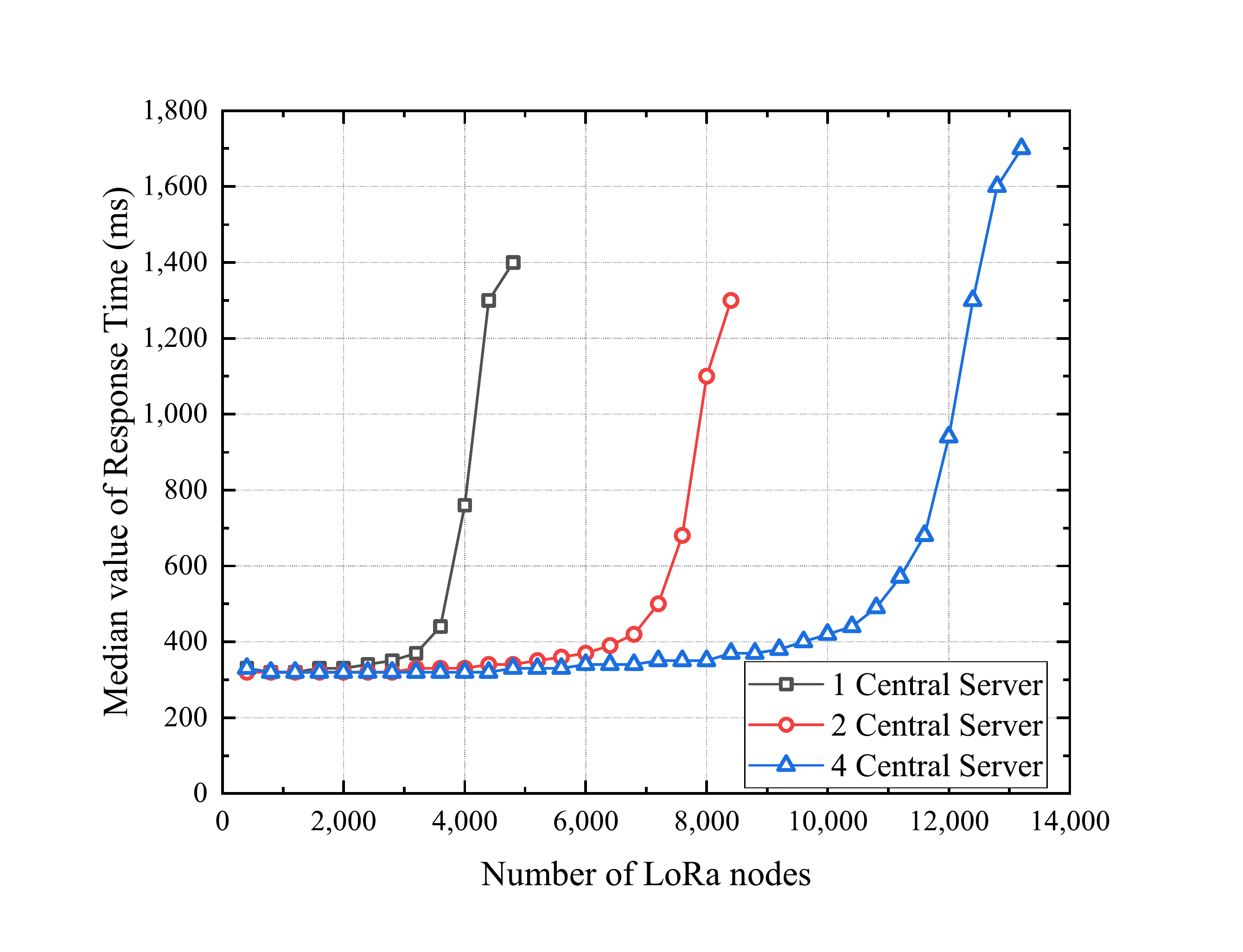}
	\caption{The median value of response time.}
	\label{response_time}
	\vspace{-1em}
\end{figure}

\begin{figure}[!t]
	\centering
	\includegraphics[width=0.53\textwidth]{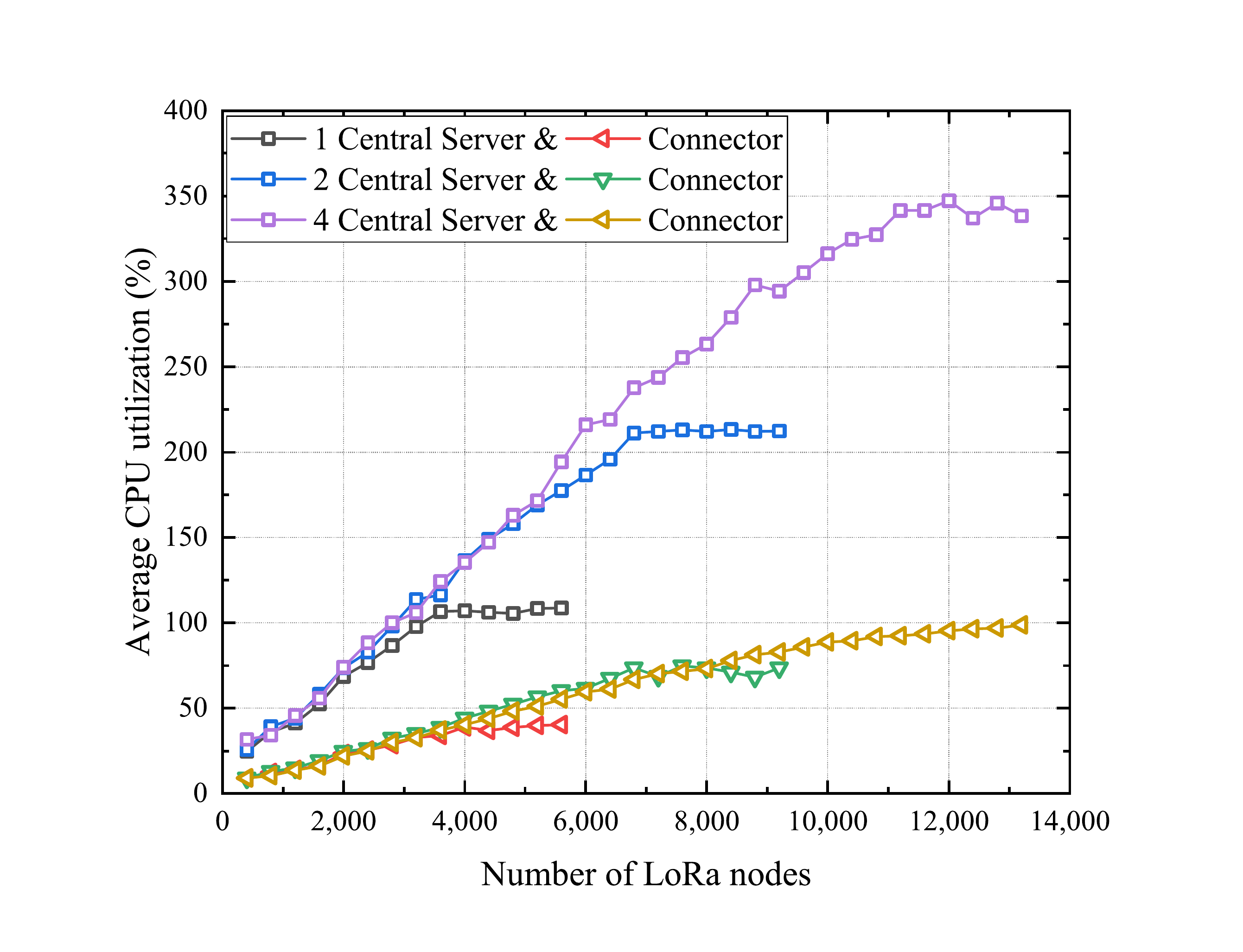}
	\caption{Average CPU utilization of Connector and Central Server.}
	\label{CPU_utilization}
	\vspace{-1em}
\end{figure}
As shown in Fig.\ref{throughput}, the throughput of LoRa network server is linear with the increasing number of LoRa Nodes in the region of small number since LoRa network server can successfully process all packets. However, the throughput becomes floor when the number of LoRa node reaches a specific limit, e.g., 4000, 7,600, and 12,400 for 1, 2 and 4 Central Server, respectively. Meanwhile, the maximum throughput is about 98, 190 and 300 packets per second. It is because that too many packages cause the network server unable to successfully process packets in time. Figure \ref{response_time} shows that the median value of response time changes slowly when the number of Nodes is less than the limit but increases sharply when the number becomes too larger.

Figure \ref{CPU_utilization} shows the average CPU utilization of Central Server and Connector for different deployment number. It can be seen that, the CPU utilization of Central Server can slightly exceed 100\% and 200\% with 1 or 2 Central Server and the curves no longer rise with the increase of LoRa Nodes. In addition, when the CPU utilization of Connector reaches 100\%, the CPU utilization of Central Server also becomes floor. It is because these modules are a single-threaded program developed by Node.js. One Connector or Central Server can only occupy one logical CPU. When any of the logic CPU is fully occupied no matter on which module is running, LoRa network server cannot accommodate more LoRa Nodes. On the other hand, Central Server needs more computing resources than Connector so that it is more likely to become a bottleneck. Luckily, our proposed architecture supports deploy multiple modules distributedly and achieves load balancing across them.

\section{Conclusion}
Due to wide area coverage and low power consumption of LoRa network, various applications provided by LoRa have been emerging in the IoT market. In this paper, aiming at providing a flexible and completed solution for building a private LoRa network, the design and implementation of LoRa network have been proposed including hardware and software. Moreover, the open source project is available on GitHub. The field trails demonstrate that the maximum transmission distance based on our hardware design is about 7.5 $km$ in urban environments. Furthermore, we also show that LoRa system under the improved architecture can support more than 10,000 LoRa Nodes with well-deployed computing resources. It is expected that this open LoRa network can provide the flexibility and feasibility for both academic and industry to deploy their new LPWA applications in the near future.

\section*{Acknowledgment}
The work was supported by the National Natural Science Foundation of China (NSFC) under the Grant Number 61671089.

\bibliography{reference}
\bibliographystyle{ieeetr}

\end{document}